\documentstyle[psfig,amssymb]{mn}
\title[The pairing fraction of faint radio sources] 
{The Phoenix survey: the pairing fraction of faint radio sources}

\author[A. E. Georgakakis et al.] {A. E. Georgakakis$^{1,2}$, B.
  Mobasher$^{2}$, L. Cram$^{3}$, A. Hopkins$^{4}$ \\ \\
  $^1$ School of Physics and Astronomy, University of Birmingham,
  Edgbaston, B15 2TT, UK\\
  $^2$Astrophysics Group, Blackett Laboratory, Imperial College, Prince
  Consort Rd , London SW7 2BZ, UK\\
  $^3$ Astrophysics Department, School of Physics, University of Sydney,
  NSW, Australia 2006\\ 
  $^4$ Department of Physics and Astronomy, University of Pittsburgh, 3941
  O'Hara Street, PA 15260, USA\\
}
\begin{document}
\maketitle  

\begin{abstract}
The significance of tidal interactions in the evolution of the faint
radio population (sub-mJy) is studied using a deep and homogeneous radio
survey (1.4\,GHz),  covering an area of 3.14\,deg$^{2}$ and complete to a
flux density of 0.4\,mJy. Optical photometric and spectroscopic data are
also available for this  sample. A statistical approach is employed to  
identify candidate physical associations between radio sources and
optically selected `field' galaxies. We find an excess of close  pairs
around optically identified faint radio sources, albeit at a low
significance level, implying that the pairing fraction of the sub-mJy radio
sources is similar to that of `field' galaxies (at the same magnitude
limit) but higher than that of local galaxies. 
\end{abstract} 
 

\section{Introduction}\label{sec_6.1}

Visual inspection of faint radio sources shows that many have optical
counterparts which are preferentially located in pairs or small groups
exhibiting disturbed optical morphologies, suggestive of interactions or
mergers (Kron {et al.} 1985; Gruppioni et al. 1999).  Although
this is not an ideal method for identifying physically associated groups,
the frequency of cases suggests that such phenomena are at least partially
responsible for the enhanced star formation rate seen in these objects
(Kron {et al.} 1985; Windhorst {et al.} 1995). There is also independent
evidence  that tidal interactions affect both the nuclear galaxy activity
and the disk star-formation rate. For example spectroscopic studies of
interacting systems show significant  excess of nuclear optical-line
emission (e.g. H$\alpha$) compared to `field' galaxies (Keel {et al.} 1985),
due to either nuclear star-formation or a non-thermal central source. A
systematic enhancement is also seen in the disk H$\alpha$ and far-infrared
emission (e.g. Kennicutt  {et al.} 1987; Bushouse {et al.} 1988),
attributed to increased star-formation in the galactic disk. 
Similarly, galaxy interactions are shown to enhance the radio
emission from the nucleus of galaxies, which is mainly caused by star
formation activity within the nuclear region (e.g. Hummel et
al. 1990).    Recently, Gallimore \& Keel  (1993) demonstrated that about
$30\%$ of infrared (60\,$\mu$m)  selected galaxies are found in pairs
showing an increase in their pairing fraction with increasing
60\,$\mu$m luminosity. The pairing fraction of infrared selected galaxies
is comparable to that derived for optically selected starbursts (Keel \&
van Soest 1992). Moreover, a similar study of a redshift-limited sample of 
Seyfert-type galaxies (Dahari 1984) has revealed an increased fraction of
close companions ($\approx15\%$) compared  to a control sample of  `field'
galaxies.    

Galaxy interactions and mergers can be studied, in the absence of redshift 
information, using a statistical approach. This is to estimate the
probability  that galaxies, with angular separation $\theta$, are
physically associated, rather than randomly aligned on the sky plane. This
technique has been applied to deep magnitude limited surveys, to study the
evolution in the rate of  galaxy mergers with redshift (Zepf \& Koo 1989;
Burkey {et  al.} 1994; Roche {et al.} 1998). In this paper we employ this
method to estimate the pairing  fraction of the sub-mJy sources detected in
a deep and homogeneous radio survey (Phoenix) with available
multi-wavelength data. This technique has the advantage of being more
quantitative and  objective in identifying candidate interacting systems,
compared to simple visual inspection. The latter method  is sensitive to
the depth of the survey, the wavelength of the observation  and most
crucially  to the observer's criteria for identifying potentially
interacting galaxies. 

The observations  are described in section \ref{sec_2}. The
statistical method for identifying candidate interacting galaxies is
discussed in section \ref{sec_3}, while section \ref{sec_4} estimates the
minimum angular separation for which close pairs can be resolved. The
results are presented in section  \ref{sec_5} and discussed in section
\ref{sec_6}. Finally, section \ref{sec_7} summarises our conclusions.    
Throughout this paper we assume a value
$H_{0}=50\,{\mathrm km\,s^{-1}\,Mpc^{-1}}$ and $q_{0}=0.5$.

\section{Observations}\label{sec_2}
\subsection{Radio observations}
The radio observations were made at 1.4\,GHz using the 6A configuration of
the Australia Telescope Compact Array (ATCA). The mosaic of 30 pointing
centres covers a $2^{\circ}$ diameter area centred at 
 ${\mathrm RA}(2000)=01^{\mathrm h}~14^{\mathrm m}~12\fs16$; 
 ${\mathrm Dec.}(2000)=-45^{\circ}~44'~8\farcs0$. 
The synthesised beam FWHM for each of the pointing centres is
$\approx$8\,arcsec.   

Details of the observations, image formation, source extraction and
catalogue generation are presented in Hopkins et al. (1998).  A
source is included in the catalogue if its peak flux density 
is 4$\sigma$ above the local RMS noise. A total of 1079 sources with flux
densities $S_{1.4}>0.1$\,mJy are  detected. There are two kinds of
incompleteness in the catalogue, as with any sample limited by peak flux
density. The first is a loss of sensitivity due to the attenuation of the
primary beam away from a pointing centre. This has been minimised in the
Phoenix survey by the mosaicing strategy used. The second is the fact that
extended objects with a total flux density above the survey limit can be
missed by an algorithm which initially detects candidates based on their
peak flux density. Methods of correcting these effects have been
described by Hopkins et al. (1998). The radio catalogue is found to
be $\approx80\%$ complete to 0.4\,mJy. 

\subsection{Optical Photometric Observations}

The optical survey of the Phoenix field was carried out at the
Anglo-Australian Telescope (AAT) in the $R$-band. Details about the
observations and data reduction are presented in Georgakakis et 
al. (1999). The source extraction and photometry is performed using the 
 FOCAS package (Jarvis \& Tyson 1981). The star-galaxy separation is
carried out to the limiting magnitude $R=20.0$\,mag. At fainter magnitudes
no attempt is made to further eliminate stars form the sample, since
compact galaxies could be mistakenly removed (Georgakakis et al. 1999). The
resulting galaxy catalogue, complete to $R=22.5$\,mag, is used to optically
identify the sources detected  in the radio survey as described in
Georgakakis et al. (1999). A total of 504 radio sources ($47\%$)  have been 
identified to  $R$=22.5\,mag.

\subsection{Optical Spectroscopic}

The spectroscopic data were obtained using slit spectroscopy at the ESO
3.6\,m telescope and multi-object fibre spectroscopy at the 2 degree field 
spectroscopic facility (2dF) at the AAT. Details about these observations
are presented by Georgakakis et al. (1999). Redshifts were established for
228 out of 320 candidate optical identifications. The optical spectral
features of these sources were employed to classify them as (i)
absorption-line systems likely to be ellipticals; (ii) star-forming
galaxies; (iii) Seyfert 1 and 2 type galaxies; and (iv) unclassified
objects. The unclassified objects displayed at least one identified
emission line (allowing a redshift to be determined), but poor S/N, or
a very small number of emission lines within the observable window, or the
presence of instrumental features contaminating emission lines, prevented
us from carrying out a reliable classification.  

\section{The method}\label{sec_3}

To determine whether two galaxies are closely aligned on the sky by chance 
or are physically associated, we follow the method employed by Burkey {et
al.} (1994). Given a random distribution of unrelated galaxies on the 
sky,  the probability of a galaxy pair being chance projection is 

\begin{equation}\label{eq_6.1}
  P=\int_{0}^{\theta} 2\pi\alpha\rho\,\exp(-\pi\rho\alpha^{2})d\alpha=1-\exp(-\pi\rho\theta^{2}),
\end{equation}

\noindent where $\theta$ is the angular separation between the galaxies 
of the pair, with $\rho$ being the surface density of galaxies brighter than
$m$, the magnitude of the fainter of the two galaxies. The integrand is 
the probability of finding a galaxy in a ring of width $\delta\alpha$, assuming
Poisson statistics (Scott \& Tout 1989). Equation (\ref{eq_6.1}) has a
correction  that accounts for the normal galaxy clustering, as estimated by
the angular correlation function at relatively large angular
separations. Although this correction is not included in equation
(\ref{eq_6.1}), as will be discussed in section \ref{sec_5}, our analysis
takes into account the clustering of optically selected galaxies.
The surface density, $\rho$, in equation (\ref{eq_6.1}) is calculated by
integrating the counts to the limiting magnitude of the faintest  member of
the pair. This is a conservative approach since the  contribution of pair
members projected by chance is overestimated (Burkey {et al.} 1994).  In
equation (\ref{eq_6.1}) we assume that the minimum angular separation,
$\beta$,  over which individual galaxies can successfully be resolved, is
zero. However, $\beta$ depends on both the atmospheric seeing conditions
and the splitting  efficiency of the source extraction software and is
discussed in the next section. Therefore equation (\ref{eq_6.1}) is
corrected for the resolution effect as 
 
\begin{eqnarray}\label{eq_6.2}   
P & = & \int_{\beta}^{\theta} 2\pi\alpha\rho\,\exp(-\pi\rho\alpha^{2})d\alpha \nonumber \\ 
  & = & \exp(-\pi\rho\beta^{2})-\exp(-\pi\rho\theta^{2}). 
\end{eqnarray}

\noindent To avoid biases due to incompleteness, we consider the optically
identified radio sample with $S_{1.4}\ge0.4$\,mJy and $17.0\le R\le21.0$\,mag,
comprising a total of 206  sources. Regions  contaminated by bright
stars and vignetted corners are masked out. A probability cutoff
$P\le0.05$, provides a  sample of pairs that are least likely ($\le5\%$)
to be spurious alignments.  To check the sensitivity of our results to the
adopted probability cutoff, we  also consider the 
$P\le0.1$ case.    

\section{Simulations}\label{sec_4}

The atmospheric seeing conditions at the time of the observation limit the 
minimum separation for which the independent components of galaxy pairs are 
successfully detected. For the present observations the average seeing is
$\approx1$\,arcsec. Although FOCAS can, in principle, split close pairs
to the seeing limit, there is a trade-off between the FOCAS resolving
power and the number of spurious detections due to multiple splitting
of extended sources. The FOCAS splitting parameters are therefore, chosen
to minimise this effect, albeit in the expense of resolving power. To
quantify the FOCAS splitting efficiency we use IRAF software to construct
artificial galaxy images separated by  angular distances between
2--6\,arcsec and magnitudes  in the range $R=17.0-21.0$\,mag. Then, we
attempt to recover the individual  members  of the pair using the FOCAS
package as with the real data-set.  

The simulated galaxies are assigned an exponential intensity profile
(Freeman 1970) similar to that of spiral galaxies and absolute magnitudes
in the range M$_{B}$=--21.0 to --19.0\,mag, bracketing the characteristic 
luminosity (M$_{B}^{*}\approx-20.0$\,mag; Metcalfe et al. 1991) of the
galaxy luminosity function. The  peak surface density is 
$\mu_{o,B}=21.5$\,mag\,arcsec$^{-2}$, similar to  that of high surface
brightness galaxies (Freeman 1970). The transformation from the $B$ filter
to the $R$-band is performed using the $B-R$ colour at $z=0$ of a model
spectral energy distribution corresponding to Sab/Sbc type galaxies
(Pozzeti Bruzual \& Zamorani 1996; Georgakakis et al. 1999). 

\begin{figure} 
\centerline{\psfig{figure=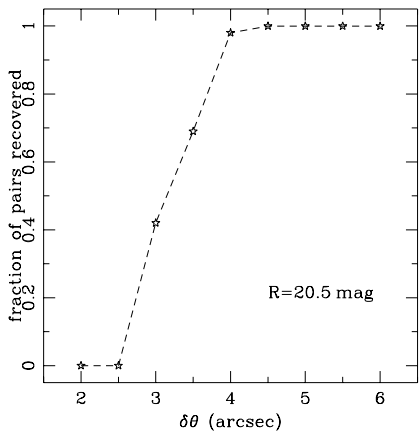,width=0.4\textwidth,angle=0}} 
 \caption
 {Percentage of successfully recovered pairs as a function of 
 the angular separation of the components. The individual galaxies 
 in this example have $R=20.5$\,mag.
 }\label{fig_6.1}
\end{figure}

For a given $M_{R}$, galaxy pairs are generated with each of the galaxy
components  having magnitudes in the range $R=17.0-21.0$\,mag. Poisson
noise was added to the image, which is then convolved  with a Gaussian
filter with FWHM of 1\,arcsec, simulating the effect of seeing. The
simulations show a  minimum resolving separation of
$\approx4$\,arcsec. This is demonstrated  in Figure \ref{fig_6.1} where the
fraction of successfully split pairs is plotted against the separation of
the components. For smaller separations FOCAS is unable to split the
individual  galaxies, implying $\beta=4$\,arcsec in equation
(\ref{eq_6.2}). Adopting $\beta=1$\,arcsec  slightly decreases the
estimated number of pairs with $\delta\theta\ge4$\,arcsec, but does not
alter any of our final conclusions.   

\section{Galaxy pair counting}\label{sec_5}

\begin{figure*} 
\centerline{\psfig{figure=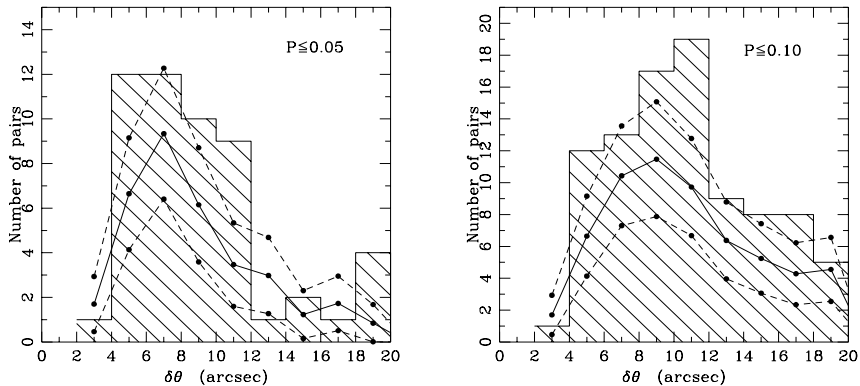,width=5in,height=2.5in,angle=270}} 
\caption
 {Number of radio-galaxy/`field'-galaxy pairs with probabilities
 $P\le0.05$ (left panel) and $P\le0.10$ (right panel) of being
 spurious alignments. The continuous line is the mean pair count
 derived for `field' galaxies.
 The dashed lines are the 1$\sigma$ envelope lines around the 
 mean.}\label{fig_6.3}
\end{figure*}

The technique outlined  in  section \ref{sec_3} is used to calculate
the number of close pairs between radio sources and `normal' optically
selected galaxies (radio-galaxy/`field'-galaxy pairs). The distribution of
radio-galaxy/`field'-galaxy pairs as a function of angular separation is
shown by the hatched histogram in Figure \ref{fig_6.3} for the probability
cutoffs $P<0.05$ and 0.10 respectively. 
However, a fraction of the identified pairs are expected to be random
superpositions on the sky plane. Moreover,  because of the normal
clustering of galaxies, quantified by the two point correlation function,
$w(\theta)$, we also expect a number of non-random galaxy pairs around
optically identified radio sources. 
Therefore, to investigate whether the pairing fraction of optically
identified sub-mJy radio sources differs from that expected for `normal'
optically  selected galaxies on average, we need to assess the significance
of these two effects. For that purpose a total of 200 mock catalogues are
constructed by randomly selecting galaxies in the range
$17.0<R<21.0$\,mag from the optical galaxy catalogue. Each of the mock
catalogues  has the same number of objects and the same magnitude
distribution as the optically identified radio sample with
$S_{1.4}\ge0.4$\,mJy 
and $17.0<R<21.0$\,mag. Taking  the galaxies of the mock catalogue in
question as centres, the number of pairs with `field' galaxies is calculated,
using the method outlined in section \ref{sec_3}. The mean number of close
companions and the standard deviation at each angular separation is
calculated from the 200 mock catalogues.  
The results are shown with the continuous line in Figure 2. The dashed
lines represent the $1\sigma$ deviations around the mean. To improve the
statistics we consider pairs in the range 4--12\,arcsec, where our estimator
is more sensitive in identifying candidate interacting systems (see section
\ref{sec_6}).    From the mock catalogues we count a total of 25.6$\pm$5.0
pairs around `normal' optically selected galaxies for angular separations 
$4<\delta\theta< 12$\,arcsec and for $P\le0.05$. This compares to
$43.0\pm6.6$ (Poisson statistics) pairs around radio sources. Therefore,
the number of pairs around faint radio sources above the random and
$w(\theta)$ expectation is $(43.0\pm6.6)-(25.6\pm5.0)=17.4\pm8.3$
(2.1$\sigma$ confidence level). A similar  result, with a slightly larger
significance (2.3$\sigma$), is obtained for $P\le0.10$.   

\section{Discussion}\label{sec_6}

The environment of faint radio sources has been explored in previous
studies using visual inspection. Kron et al. (1985) considered a  
radio selected sample with $S_{1.4}>0.6$\,mJy and available four-band
photometry ($U$, $J$, $F$, $N$-bands). For the sub-sample with
$17.0<R<21.0$\,mag (here we assume the transformation $R=F-0.14$ between
$F$ and $R$-band magnitudes; Metcalfe et al. 1991) they  found that  about
$19\%$ of the sources (18 out of 95) lie in pairs or groups. More
recently, Gruppioni et al. (1999) studied a radio sample with
$S_{1.4}>0.2$\,mJy and multi-wavelength photometric data. They conclude
that about $27\%$ of the sources with $S_{1.4}>0.4$\,mJy and
$17.0<R<21.0$\,mag (6 out of 22 galaxies in that study) are found in pairs
or small groups.  The analysis described in the previous section
identified 43 candidate interacting pairs out  of a total of 206
sources, corresponding to a pairing fraction of $21\%$, in reasonable
agreement with the results from the above mentioned studies. 

We also find that although the pairing fraction of the sub-mJy sources in
the present sample ($S_{1.4}\ge0.4$\,mJy; $17.0\le R \le 21.0$\,mag;
$\delta\theta\ge4$\,arcsec) is higher than that of optically selected
galaxies (at the same magnitude limit), the significance of the 
excess is low ($\approx2\sigma$). 
This implies that the pairing fraction of the sub-mJy population is higher
than that of local galaxies. This is because the frequency of interactions
in optically selected samples has been shown to increase with redshift
(Burkey et al. 1994). Therefore, the optically selected galaxy sample
studied here, with a magnitude limit $R=21.0$\,mag, corresponding to a
mean redshift $z\approx0.25$ (Pozzetti, Bruzual \& Zamorani 1996), already
comprises a higher fraction of interacting systems than local galaxy
samples. 

Moreover, the similarity of the sub-mJy and `field' galaxy close pair
distributions in Figure \ref{fig_6.3} suggests that radio selection at a
given optical magnitude limit does not guarantee a significantly higher
fraction of interacting systems compared to optical selection (at the same
limiting magnitude).  

However, one should be cautious about this interpretation.
Firstly, the statistical approach used here calculates the probability of a
galaxy having a physically associated companion with projected separation
$\delta\theta$ and magnitude $R$ by estimating the expected number of
galaxies brighter than $R$ within radius $\delta\theta$. Therefore, for a
given magnitude, $R$, the larger the separation, $\delta\theta$, the higher
the probability, $P$, that the pair is a spurious alignment of the sky
plane. For example, for $R\approx20.0$\,mag only pairs with
$\delta\theta\lesssim12$\,arcsec are identified as potentially interacting 
for $P<0.05$.  Consequently, the estimator in equation (\ref{eq_6.2}) is
sensitive to relatively close  pairs ($\delta\theta\lesssim12$\,arcsec) and
is likely to miss physically associated galaxies with larger projected
separations.  
Additionally, due to resolution effects we can only identify candidate
interacting systems with $\delta\theta\gtrsim4$\,arcsec. This angular
separation corresponds  to a linear size of $20\,\mathrm{kpc}$
($H_{0}=50\mathrm{\,km\,s^{-1}\,Mpc^{-1}}$) at $z=0.25$, the median
redshift of the sub-sample with  $S_{1.4}\ge0.4$\,mJy, $17.0\le R\le21.0$\,mag 
and available spectroscopic information (79 sources out of a 
total of 206). These very close pairs are also expected to have high radio
luminosity (normalised to optical luminosity; Read \&  
Ponman 1998). Future high-resolution photometric observations  have the
potential to reveal the presence of very close pairs within the sub-mJy
population, that remain unresolved in the present study.    

Moreover, in our analysis we only consider the optically brightest
($R\le21.0$\,mag) radio sources. These are likely to lie, on
average, at relatively lower redshifts compared to optically fainter
objects ($R>21.0$\,mag). The pairing fraction of the optically fainter
sub-mJy sources remains to be explored.

\begin{figure} 
\centerline{\psfig{figure=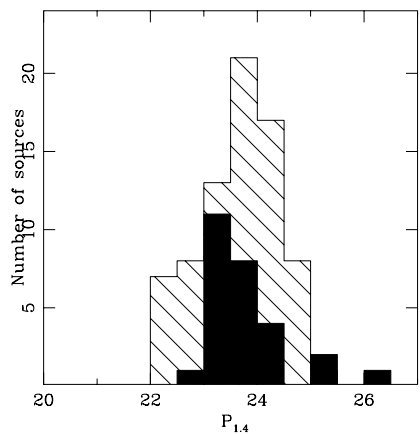,width=0.4\textwidth,angle=0}} 
 \caption
 {Radio power distributions for `isolated' (hatched region) and
 paired (shaded histogram) radio sources (see text for
 details).}\label{fig_6.4} 
\end{figure}

\begin{figure} 
\centerline{\psfig{figure=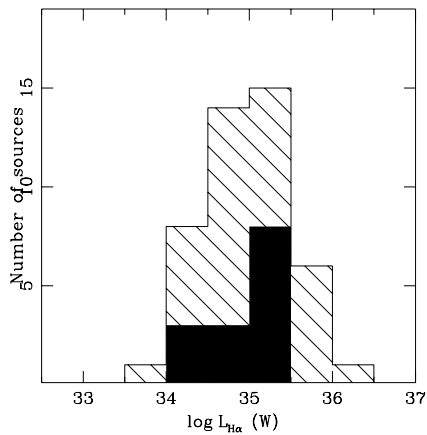,width=0.4\textwidth,angle=0}} 
\caption
 {H$\alpha$ luminosity distribution for isolated (hatched region) 
 and paired (shaded histogram) radio sources.}\label{fig_6.5}
\end{figure}

A total of 27 out of 43 ($\approx60\%$) faint radio sources for which our
analysis indicates possible association with `field' galaxies ($P\le0.05$),
have available spectroscopic information.  The sample  consists of
absorption-line systems ($37\%$), Seyfert 1 or 2 type objects ($15\%$),
unclassified sources ($30\%$) and star-forming galaxies ($19\%$). For
$P\le0.10$, the relative fraction of different types of sources is
similar. 
The radio power distributions of these sources are compared in Figure
\ref{fig_6.4} with those of the spectroscopic sample of 
optically identified radio galaxies with $S_{1.4}\ge0.4$\,mJy and
$17.0\le R\le 21.0$\,mag. The  histogram for the H$\alpha$ luminosity
distribution of the two samples is shown in Figure  \ref{fig_6.5}. There is
no significant difference between the two  populations in these
Figures. Hummel (1981) also found no difference in the radio power
distributions of isolated galaxies and galaxies in pairs.
Moreover, interacting systems, although being on average more luminous at
infrared or H$\alpha$ wavelengths compared to `field' galaxies,
exhibit significant scatter around their mean emission properties (Keel et
al. 1985; Kennicutt et al. 1987; Bushouse et al. 1988). This is
attributed to (i) the relative velocities of the interacting galaxies  (ii)
the properties of the galactic disks, especially in relative weak
encounters and (iii) the time scale of the induced star-formation relative
to that of the gravitational encounter (Kennicutt et al. 1987). Hence,  a
relatively small sample like the one considered here, cannot reveal any
correlations between pairing fraction and optical emission-line
luminosity. Moreover, close pairs with projected separations
$\delta\theta<4$\,arcsec remain unidentified in the present study. These
pairs are also likely to be on average more luminous at optical and
infrared wavelengths compared to systems with larger separations (Read \&
Ponman 1998).   

Additionally, our analysis shows that galaxies with non-thermal
nuclear activity are more frequently found in associations ($\approx50\%$)
compared to radio sources heated by hot stars. This is because there is
only a small number of star-forming galaxies  ($\approx20\%$ of the
spectroscopic sample; Georgakakis et al. 1999) in the sub-sample with
$S_{1.4}\ge0.4$\,mJy. This is dominated by absorption-line systems, Seyfert
1 and 2s, constituting   $\approx64\%$ of the $S_{1.4}\ge0.4$\,mJy
spectroscopic sample (Georgakakis et al. 1999).  

\section{Conclusions}\label{sec_7}

In this study an objective and quantifiable statistical approach is
employed to assess the significance of tidal interactions in the
evolution of the faint radio sources, detected in a deep and
homogeneous radio survey with available photometric and spectroscopic
data. In particular, the pairing fraction of the faint radio
population is compared with that of optically selected `field'
galaxies. We found evidence for an excess of close pairs around 
optically identified faint radio sources, albeit at a low significance
level. This implies that (i) the frequency of interacting systems within
the sub-mJy  population is higher than that in local galaxy samples and
(ii) the pairing properties of the sub-mJy radio sample (with the given
biases such as resolution effects) are not significantly  different from
those of `field' galaxies at the same magnitude limit. 

\section{Acknowledgements}

During part of this study AG was supported by the State Scholarship
Foundation of Greece (IKY). We thank the referee for helpful comments and
suggestions that improved this paper.

\end{document}